\documentclass[12pt]{iopart}
\usepackage{epsfig}

\newcommand{\bra}{{\langle}}
\newcommand{\ket}{{\rangle}}
\newcommand{\eg}{{\it e.g.,}\ }
\newcommand{\ie}{{\it i.e.,}\ }
\newcommand{\ls}{\ell_s}
\newcommand{\reef}[1]{(\ref{#1})}

\newcommand{\myfig}[3]{\begin{figure} %[ht]
\begin{center}
\leavevmode \epsfxsize=#2cm \epsfbox{#1}
\end{center}
\caption{#3} \label{fig:#1}
\end{figure}}

\begin{document}

\title[Quark Soup al dente]{Quark Soup al dente: Applied Superstring Theory}

\author{R. C. Myers$^{1,2}$ and S. E. V\'azquez$^1$}

\address{$^1$ Perimeter Institute for Theoretical Physics, 31 Caroline St. N., Waterloo,
Ontario N2L 2Y5, Canada\\
$^2$ Department of Physics and Astronomy, University of Waterloo,
Waterloo, Ontario\\
\ \ N2L 3G1, Canada}

\ead{rmyers@perimeterinstitute.ca, svazquez@perimeterinstitute.ca}
\begin{abstract}
In recent years, experiments have discovered an exotic new state of
matter known as the strongly coupled quark-gluon plasma (sQGP). At
present, it seems that standard theoretical tools, such as
perturbation theory and lattice gauge theory, are poorly suited to
understand this new phase. However, recent progress in superstring
theory has provided us with a theoretical laboratory for studying
very similar systems of strongly interacting hot non-abelian
plasmas. This surprising new perspective extracts the fluid
properties of the sQGP from physical processes in a black hole
spacetime. Hence we may find the answers to difficult particle
physics questions about the sQGP from straightforward calculations
in classical general relativity.

\end{abstract}

\maketitle

\section{Introduction}
Quantum chromodynamics (QCD) is the theory of the ``strong force"
which determines the physical properties of protons, neutrons and
more generally hadrons. At a superficial level, QCD looks like a
matrix-valued version of the more familiar electromagnetic theory,
replacing photons by gluons and electrons by quarks. However, unlike
electromagnetism, quantum fluctuations of the fields play an
essential role in determining the force law in QCD. In particular,
at low energies or large distances (by the standards of subatomic
physics), the coupling of QCD is large and the forces are ``strong",
as implied by the original name. This is the origin of confinement,
\ie the fact that we do not see free quarks or gluons in nature but
rather we only see ``colourless" or QCD-neutral packages known as
hadrons. However, at high energies or short distances, the QCD
coupling is small and correspondingly the forces are weak. Known as
asymptotic freedom, this property allows us to detect the composite
structure of hadrons by, \eg scattering high energy electrons.

Because of the strong coupling, our understanding of many aspects of
QCD remains incomplete. For example, while it has now been more than
thirty-five years since the discovery of asymptotic freedom, a
complete theoretical understanding of confinement remains elusive.
In fact, an analytical proof of confinement is now  one of the Clay
Institute's ``Millennium Problems" for which there is a one-million
dollar prize \cite{millennium}. Of course, great progress has been
made on various theoretical fronts. For example, lattice gauge
theory \cite{lattice}, which essentially puts QCD on a computer, now
provides a fairly good description of the physical properties of low
energy hadrons \cite{hadron}. Another approach is to study QCD away
from its ground state. In particular, asymptotic freedom indicates
interactions are weaker at short distances and high energies and so
one might expect to find new behaviour for QCD at high densities and
high temperatures. In fact, a combination of lattice and analytic
efforts have allowed theorists to map out the phase diagram of QCD
\cite{phased} illustrated in figure \ref{fig:phase1}.
\myfig{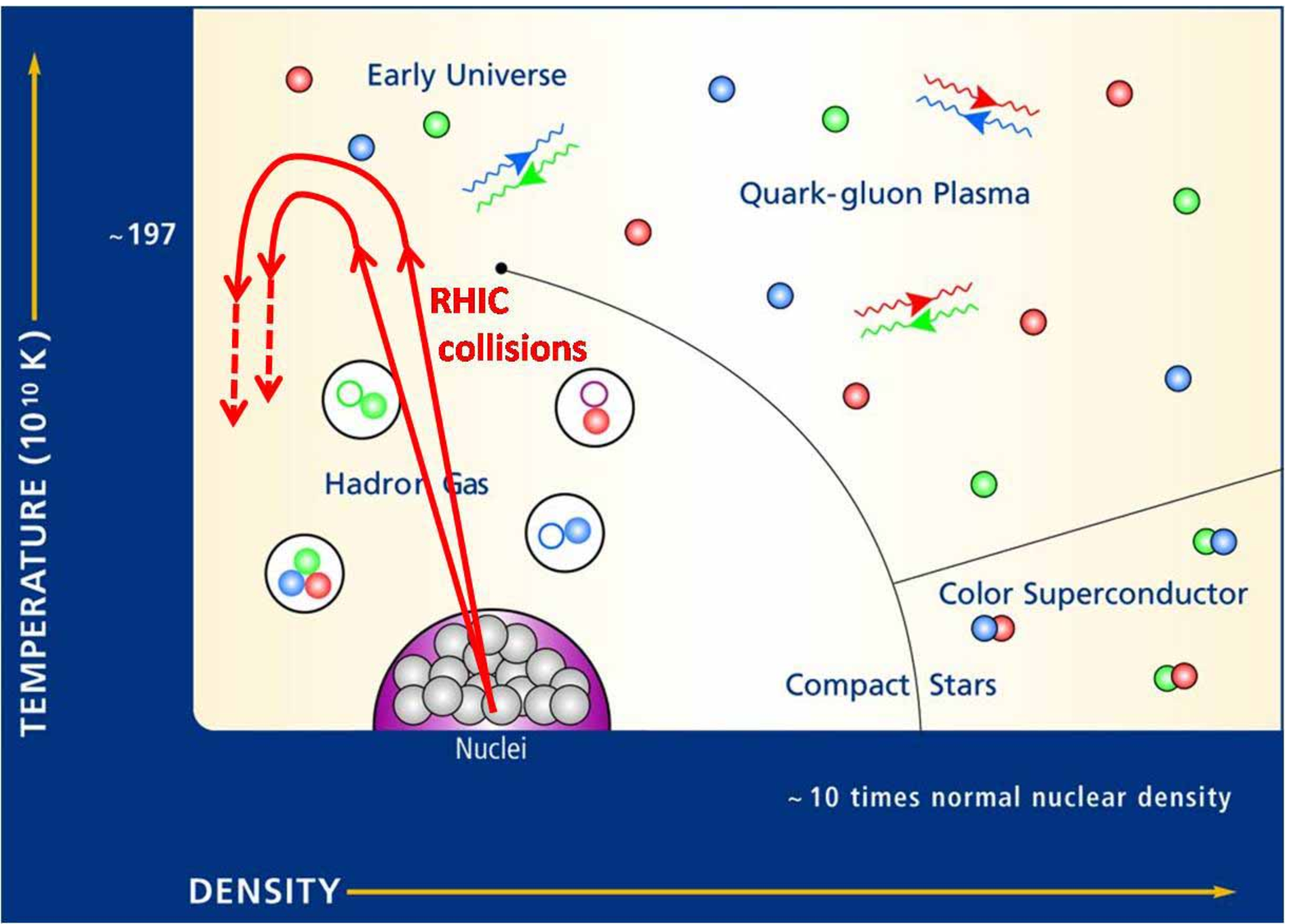}{14}{Phase diagram of QCD according to theorists
\cite{phase}. The red arrows show caricatures of the evolution of
matter in typical RHIC collisions.}

Of course, we should also seek an experimental verification of this
theoretical picture and this was the goal behind constructing the
Relativistic Heavy Ion Collider (RHIC) at the Brookhaven National
Laboratory on Long Island \cite{RHIC}. In fact, experiments there
have revealed a new phase of hadronic matter called the strongly
interacting Quark-Gluon Plasma (sQGP). In this new phase, the quarks
and gluons are neither confined nor free but instead, form some kind
of strongly interacting ``soup" which seems to behave like a near
perfect fluid. The sQGP is created at RHIC by colliding two gold
nuclei at $\sim 200$ GeV per nucleon. The collision creates a hot
plasma of quarks and gluons which expands outward in a collective
flow. The plasma eventually cools down to a temperature where the
particles in the plasma are again confined into hadrons which then
escape out into the detectors. The red arrows in figure
\ref{fig:phase1} show typical trajectories in the phase diagram
which might describe this process. The sQGP would be probed on the
upper reaches of these trajectories. Much of this evolution was
actually predicted theoretically some time ago in \cite{Bjorken},
however, a precise interpretation of these experiments now calls for
understanding both strong coupling and also dynamical properties of
QCD. Hence we face a challenge since few if any effective
theoretical tools exist for this purpose.

At this point, the reader may well be wondering what any of this has
to do with gravity, which is the main topic of these proceedings.
Quite surprisingly, it turns out a great deal! In a parallel set of
developments, string theorists have been uncovering deep connections
between strongly coupled gauge theories and gravity. The best
understood example of such duality is the so-called AdS/CFT
correspondence \cite{malda, adscft} --- for reviews see
\cite{review1}. These dualities realize a holographic description of
quantum gravity in which the theory has an equivalent formulation in
terms of a nongravitational theory on the boundary of the original
spacetime \cite{holo}. A version of this duality involves, on one
hand, a (maximally) supersymmetric gauge theory in four dimensions
known as ${\cal N} = 4$ super-Yang-Mills theory (SYM). The duality
proposes that this gauge theory is equivalent to type IIB string
theory in an $AdS_5 \times S^5$ background. Further, in a certain
strong coupling limit, the latter essentially reduces to classical
(super)gravity in this background. In other words, one can calculate
gauge theory observables at strong coupling by doing a classical
gravity calculations.

It may seem that string theory has provided a way of solving QCD
analytically and so we should be ready to collect our million-dollar
Millenium Prize. Unfortunately, at present, the gauge theories for
which we know their gravity dual are quite different from QCD.
Nevertheless, there are regions in phase space for which SYM theory,
for example, seems to share many features in common with QCD. In
particular, this happens at high temperatures, precisely in the sQGP
phase! The aim of this article is to introduce the AdS/CFT
correspondence and explain the relations between QCD and SYM at
finite temperature. We will also briefly describe how certain
gravity calculations yield observables that are relevant for the
physics of the sQGP. Surprisingly, the values calculated using the
AdS/CFT duality turn out to be not too far from the ones observed
for QCD!

 \section{The AdS/CFT Correspondence }

String theory is much more than just a theory of strings. In
particular, it also contains non-perturbative solitonic objects
known as Dirichlet-branes or D-branes for short.\footnote{Sometimes
we use the nomenclature Dp-brane, where $p$ is some integer. This
integer reflects the number of spatial dimensions in which the
D-brane extends. For example, a D-brane with three spatial
dimensions is called a D3-brane.} Since their discovery in 1995,
D-branes have been one of the main stars of string theory
\cite{polchinski}. %,clifford}.
D-branes can be visualized as membrane-like objects where open
strings can end (see figure \ref{fig:dbrane2}).
\myfig{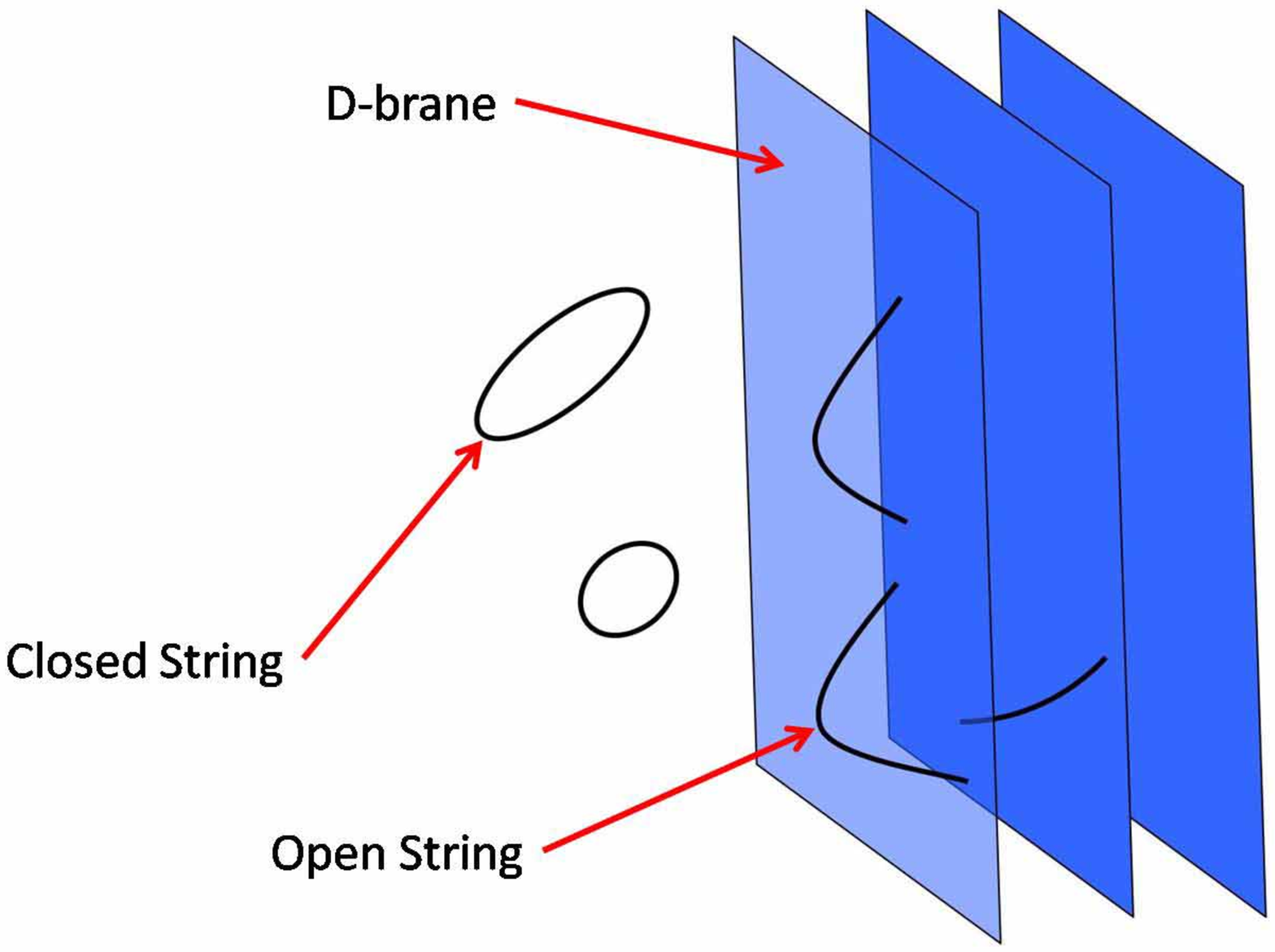}{10}{D-branes are solitonic objects of string
theory.}

Of course, string theory has both open and closed fundamental
strings, and both can interact with the D-branes. Open strings can
be visualized as excitations of the D-brane itself. Closed strings
can be absorbed or emitted from the D-brane. Since the closed string
modes include the spin-2 graviton, this absorption/emission process
is nothing but the backreaction of the D-brane on the spacetime. The
string interactions amongst themselves and with D-branes are
governed by a parameter known as the string coupling constant $g_s$.
D-branes have a tension of the order $T \sim 1/g_s$, and so at weak
string coupling they become ``heavy" solitonic objects.

D-branes have different descriptions, depending on where in
``parameter space" we are doing the calculation. For example, if we
have some finite number $N$ of D3-branes sitting on top of each
other in flat space, one finds that the low energy dynamics of the
open string excitations can be described by a non-abelian gauge
theory. This theory is precisely four-dimensional ${\cal N} = 4$ SYM
theory with gauge group $SU(N)$. The two indices of the non-abelian
fields, \eg $(A_\mu)_i^j$ with $i,j = 1, \ldots, N$, can be
visualized as labeling in which D-brane the open string is ending.
Moreover, the string coupling is related to the gauge theory
coupling by
 \begin{equation}\label{couple}
 g_{YM}^2 = 4 \pi\, g_s\;,
 \end{equation}
where $g_{YM}$ is the SYM coupling.  Note that the string coupling
$g_s$ is actually related to the expectation value of a field called
the dilaton. However, in the string theory solutions that we will
consider here, the dilaton takes a constant value and so we can tune
$g_s$ to have any value we want. This fact is also related to the
conformal symmetry of the SYM theory described above. For a string
background dual to a more complicated non-conformal theory such as
QCD, $g_s$ would no longer be a constant.

The second description of D-branes is in terms of closed strings. In
considering the backreaction of the $N$ coincident D-branes, one
finds that the size of their ``gravitational footprint" is
proportional to $g_sN$. Hence if we consider a strong coupling limit
where $g_sN \rightarrow \infty$, we cannot ignore the backreaction
on the spacetime geometry.  It is well known that the ``throat"
geometry near the D3-branes takes the form of $AdS_5 \times S^5$
\cite{review1}. In the Poincar\'e patch, the metric is
 \begin{equation}
 \label{adsmetric}
ds^2 = R^2 u^2 \left(-dt^2 + d\vec{x}^2\right)  + R^2
\frac{du^2}{u^2} + R^2 d\Omega_5^2\;.
 \end{equation}
The five-sphere part of the geometry comes from the six transverse
directions to the D3-brane in the ten-dimensional space. The radius
of curvature for both the $AdS$ and the sphere is given by,
 \begin{equation} \label{Rls}
\frac{R^4}{\ls^4} = 4 \pi g_s N = g_{YM}^2 N\;.
 \end{equation}
Here $\ls$ is the string scale which is the only free parameter of
string theory. It is related to the tension of the fundamental
strings by $T_f =  1/(2 \pi \ls^2)$. In most practical calculations,
one performs a Kaluza-Klein reduction on the $S^5$ and treats the
resulting theory as five-dimensional.
%From the five-dimensional perspective, Newton's
%constant has the value
% \begin{equation}\label{G5}
%G_5 = \frac{\pi R^3}{2 N^2}\;.
% \end{equation}

\myfig{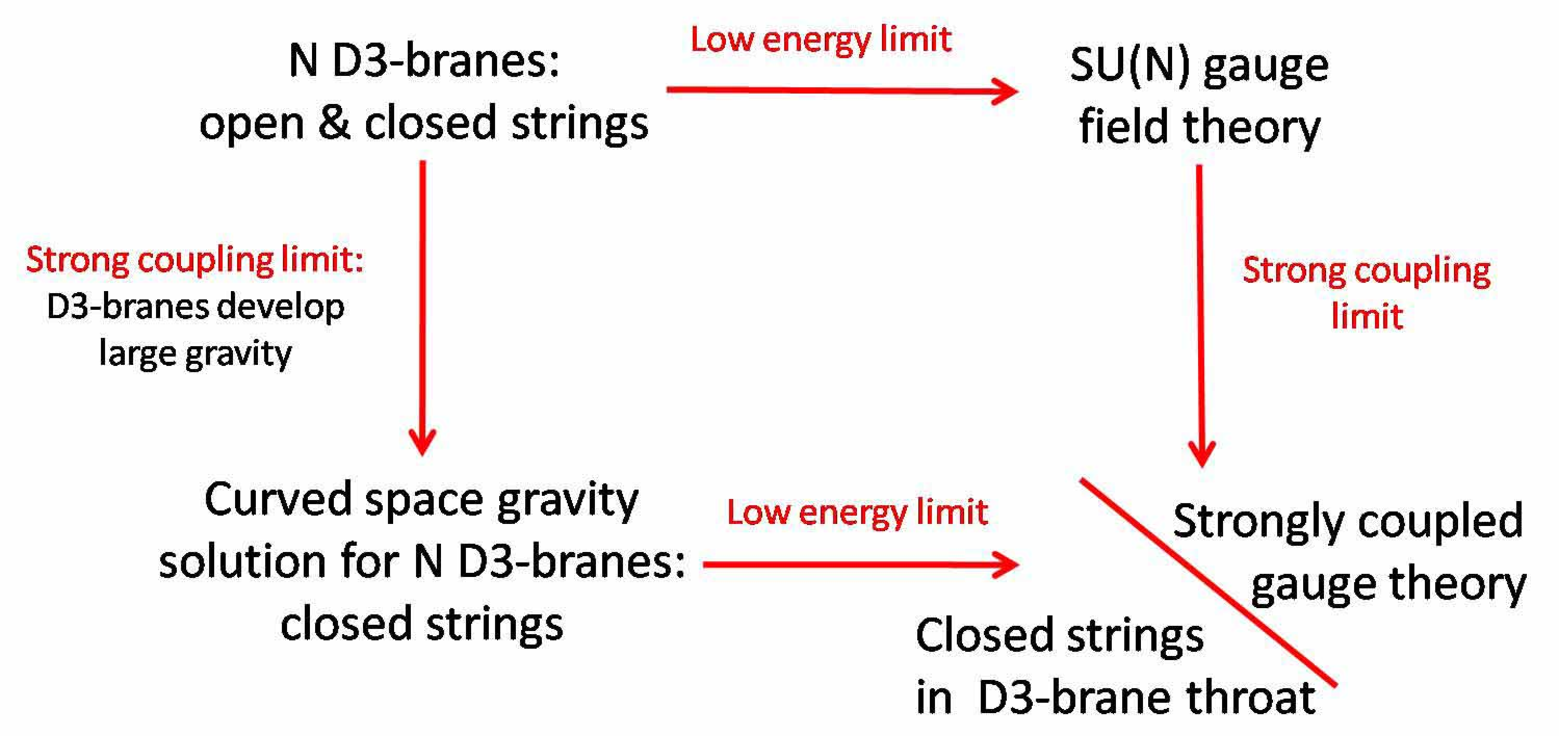}{14}{``Derivation" of the  AdS/CFT correspondence.}
In 1998, Maldacena realized that if these two limits, \ie the low
energy and strong coupling limits, were applied sequentially, two
radically different pictures emerged depending on the order of
limits, as shown in figure \ref{fig:adscft2}. However, his bold
conjecture was that these two pictures should still describe the
same physics \cite{malda}, realizing a holographic description of
the string theory. Hence the AdS/CFT correspondence proposes that
the four-dimensional ${\cal N} = 4$ $SU(N)$ SYM theory at strong
coupling should be equivalent to ten-dimensional superstring theory
on the $AdS_5 \times S^5$ background.

This correspondence can be further simplified in a particular
regime. First, if we keep the curvature scale less than the string
scale, \ie $R/{\ls} \ll 1$, then stringy corrections to the
geometric side of the duality are minimized and we may work with the
gravity theory (rather than the full string theory). Similarly if
the string coupling is kept small, \ie $g_s\ll1$ , we also reduce
the quantum or loop corrections on the geometric side and so in fact
this part of the correspondence is reduced to classical gravity.
Given the relation \reef{couple}, one might worry that the latter
limit is trivial for the gauge theory. However, comparing with
Eq.~\reef{Rls}, we see that maintaining the first inequality above
requires
 \begin{equation} \label{limit}  N \rightarrow \infty
\;,\;\;\;\;  \lambda \equiv  g_{YM}^2 N   \gg1\;.
 \end{equation}
The limit (\ref{limit}) is also known in the gauge theory as the 't
Hooft limit, where we take the rank of the gauge group $N$ to
infinity while keeping $\lambda$ fixed \cite{tHooft}. In fact, for
present purposes, the gauge theory becomes strongly coupled in
$\lambda$. Hence in the strict large $\lambda$ limit, the dual
theory reduces to just classical supergravity. This is the limit we
will be discussing in this article.

The basic dictionary between SYM theory and the gravitational theory
was established in \cite{adscft}. However, many more details of this
map have been uncovered during the last ten years. In the most basic
set up, the AdS/CFT dictionary relates every field in supergravity
(or the full string theory) to a corresponding gauge invariant
operator of SYM theory. The purpose of this lecture is not to
provide a full review of the subject. For that, we refer the reader
to \cite{review1}.  We will, however, point out the most basic entry
in the dictionary, which is the relation between the CFT partition
function and the gravitational action.

Given any number of fields in the bulk $AdS$ space (including the
metric), the partition function of the CFT is given at strong
coupling by,
 \begin{equation}\label{adsdic}
\bra \exp\left( \int \phi_\infty^i {\cal O}_i \right) \ket= \exp
\left( -I_{AdS}[\phi^i]\right)\;.
 \end{equation}
The right hand side of (\ref{adsdic}) is in reality a saddle point
approximation when the supergravity approximation is valid. The
action $I_{AdS}$ refers to the full bulk action including the
Einstein-Hilbert term evaluated as a functional of all fields that
live in $AdS$. Note that, in the saddle point approximation, this
action is evaluated on the solutions to the classical equations of
motion for all fields whose boundary condition at infinity ($u
\rightarrow \infty$ in (\ref{adsmetric})) are $\phi_\infty^i$. One
can then calculate all sorts of correlation functions of the gauge
theory by functionally differentiating on both sides of
(\ref{adsdic}) with respect to the boundary values $\phi_\infty^i$.

We will not go into the details of this procedure, which can be
quite complicated as the bulk action needs to be renormalized.
However, we will point out one result that will be useful later: how
to calculate the stress tensor of the CFT. It turns out that the
stress energy tensor of the gauge theory is dual to the metric in
the bulk. The authors of \cite{Tmunu} have given a very general
procedure to calculate such stress tensor for any metric in the bulk
that is asymptotically $AdS$.

First, we write the five-dimensional bulk metric in Fefferman-Graham
coordinates:
 \begin{equation}\label{metric2} ds^2 = R^2 \left( \frac{d\rho^2}{4 \rho^2}
  + \frac{1}{\rho} G_{\mu \nu}(x, \rho)dx^\mu dx^\nu\right)\;,
 \end{equation}
where the four-dimensional metric $G_{\mu \nu}$ has the expansion
near the boundary (now at $\rho  = 0$)
 \begin{equation}\label{asymptotic}
 G_{\mu \nu}(x,\rho) = g^{(0)}_{\mu \nu} + g^{(2)}_{\mu \nu}\,\rho+
 g^{(4)}_{\mu \nu}\,\rho^2 + \hat{g}^{(4)}_{\mu \nu}\,\rho^2 \log \rho + \ldots
 \end{equation}
The AdS/CFT dictionary in this case tells us that $g^{(0)}$ is the
metric on which the CFT is defined (flat space in most cases) and
the stress tensor is given by
%\footnote{There are additional terms in
%(\ref{T}) but they are not important if our gauge theory is defined
%in flat, or close to flat space.},
 \begin{equation} \label{T} \bra T_{\mu \nu}\ket_{CFT} =
 \frac{ N^2}{2\pi^2}\, g_{\mu \nu}^{(4)}\;.\end{equation}
We will make good use of this equation in due course.

A few comments are in order. Note that the spacetime $AdS_5$ can be
seen as the ``ground state" of the gauge theory. Any other field
that we turn on (including the metric) will result in an excited
state of the theory and/or a deformation as in (\ref{metric2}). In
the example of the stress tensor, the asymptotic metric $g^{(0)}$ is
the ``deformation", and the subleading term $g^{(4)}$ will describe
in which state the theory is. Given \reef{adsmetric}, it is easy to
show that if our space is exactly $AdS_5$ with $g^{(0)}$ the
Minkowski metric then $g^{(4)} = 0$ and so it follows that $\bra
T_{\mu \nu}\ket = 0$. Of course, this is the correct value for the
Poincar\'e-invariant ground state of SYM. In particular, if we want
to study the sQGP, we will need to study the theory at finite
temperature. As we will see in the next section, this amounts to
replacing the $AdS$ metric with a $AdS$-black hole in the bulk.

So what do we have here so far?  Well, we have uncovered a
remarkable new method to study the ${\cal N}=4$ SYM theory in the 't
Hooft limit \reef{limit}. However, in the introduction, our goal was
to study QCD. It seems that SYM is very different from QCD. In QCD,
the gauge group is $SU(3)$, \ie $N=3$ and we also have $N_f=3$
flavours of quarks, each of which are fermions transforming in the
fundamental representation.\footnote{That is, the quarks carry one
$SU(3)$ index. Recall from the above discussion, the gluons, which
transform in the adjoint representation, carry a pair of indices.}
In contrast, for SYM, we working in the limit of large $N$ and the
matter sector contains fermions and scalars, both of which transform
in the adjoint representation like the gluons. These extra fields
are required to make the SYM theory supersymmetric. However, with
this field content, the SYM theory is also a Conformal Field Theory
(CFT). That is, it is invariant under conformal transformations,
which include the usual scaling of the metric. On the other hand,
QCD is not. In fact, QCD is known to produce a dynamical scale which
is related to the confining process. Moreover the QCD coupling runs
with energy, so the string theory solution (if there is one) will
involve a running value of the dilaton which eventually becomes big
enough to invalidate the supergravity approximation.
\myfig{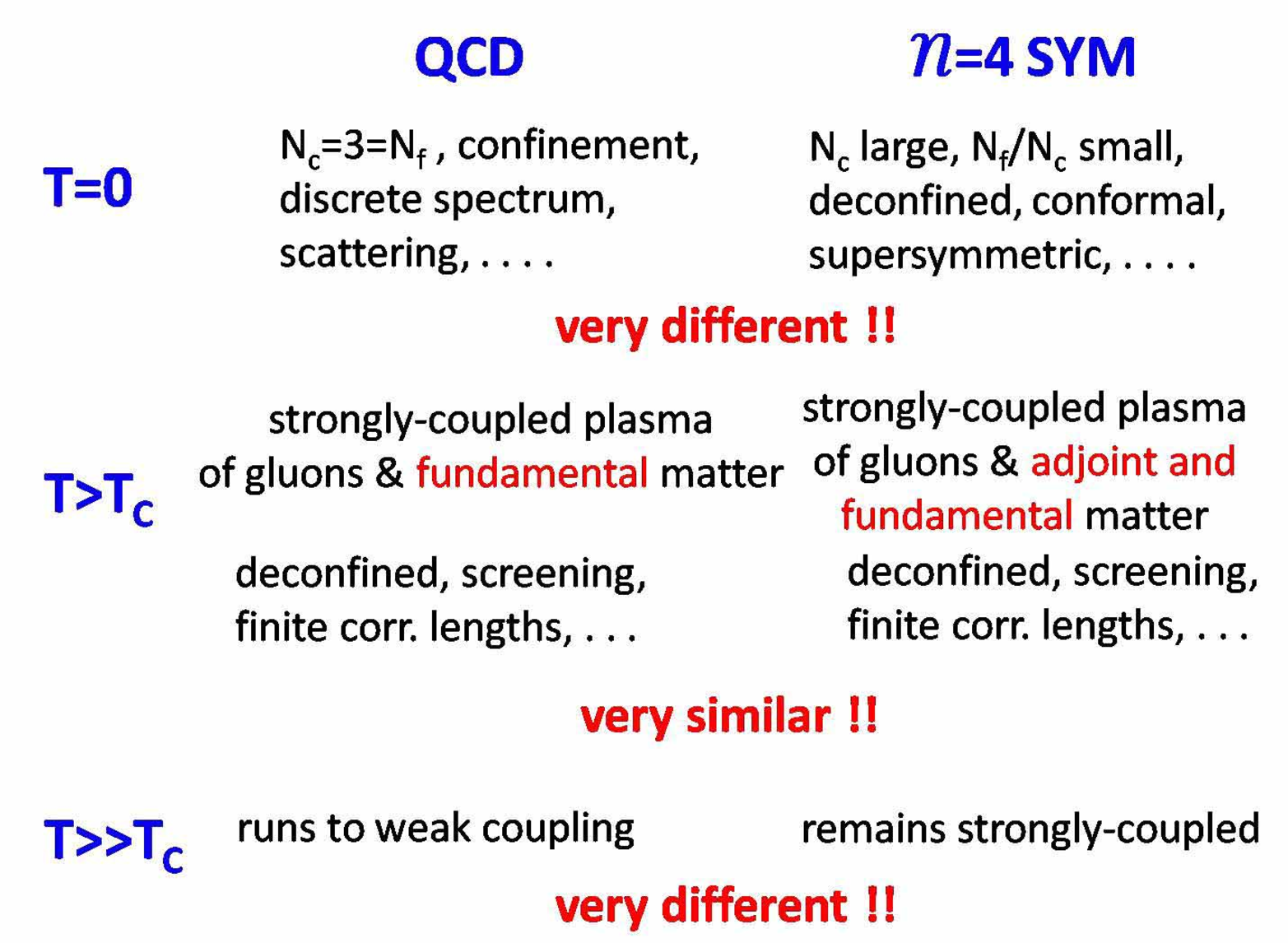}{14}{Comparison of QCD and ${\cal N} = 4$ SYM as a
function of temperature.}

Well, so far the comparison seems hopeless. However, as discussed in
the introduction, we are really interested in QCD at finite
temperature. The sQGP phase appears just above the critical
temperature where the theory becomes deconfining. For the SYM
theory, a finite temperature breaks conformal invariance and
supersymmetry by introducing a dimensionful parameter in the theory
(the temperature). Hence both systems seem to be strongly coupled
plasmas of gluons and various matter fields in this regime --- see
figure \ref{fig:compare2}. Further we emphasize that we would like
to model these plasmas with {\it fluid dynamics}, and so we only
care about long wavelength phenomena and not the microscopic
dynamics \cite{muronga}. Therefore, it is very useful to have a
strongly coupled gauge theory plasma for which we can do {\it
analytic} calculations. That is, we might hope to gain new insights
into the sQGP of QCD from our AdS/CFT calculations for SYM at finite
$T$.

\myfig{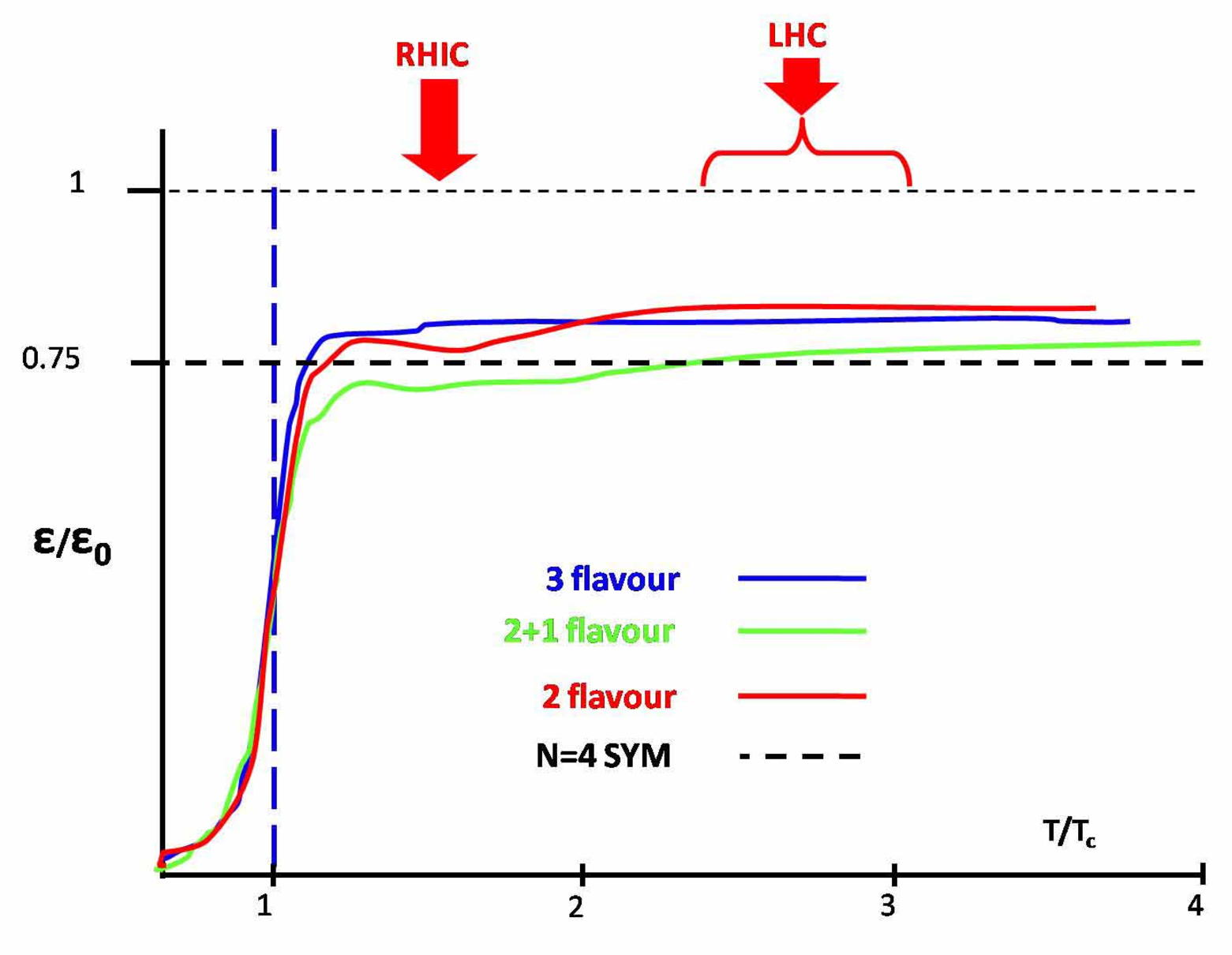}{14}{Energy density of QCD and SYM as a function of
temperature. The energy density $\varepsilon$ has been scaled by its
value at zero coupling $\varepsilon_0$. The temperature is shown in
units of the critical temperature.}
While the story above may seem very attractive, one must be
wondering if it is more than just hand-waving. In fact, there is
some evidence that the similarity between SYM and QCD at finite
temperature is more than just of a qualitative nature. One
suggestive result is shown in figure \ref{fig:lattice3}. This figure
shows lattice results for how the energy density $\varepsilon$ of
various theories close to QCD changes with the temperature
\cite{latte}. However, in this plot, the energy density has been
divided by the corresponding density when we turn off the coupling
$\varepsilon_0$. We see that when the theories enter the deconfining
regime, this ratio rises dramatically and reaches a plateau which is
close but still significantly lower than one, \ie
$\varepsilon/\varepsilon_0 \sim 0.80$-$0.85$. We might make two
observations here: First of all, despite the microscopic differences
between the various theories here, they are behaving quantitatively
in a very similar fashion when the energy density is presented in
this particular fashion. Second, the fact that
$\varepsilon/\varepsilon_0 < 1$ is an indication that on this
plateau these theories are still strongly interacting. Now figure
\ref{fig:lattice3} also shows the same ratio for the SYM theory. In
this case, the ratio is always a fixed constant which takes the
value $\varepsilon/\varepsilon_0 = 0.75$ \cite{peet}. Quite
remarkably then, despite their microscopic differences, the QCD-like
and SYM theories produce quantitative results that are very close to
each other. Hence, this lends concrete support the idea that various
aspects of the strongly coupled plasmas may be universal (or nearly
universal) and so might be amenable to study with the AdS/CFT
techniques.\footnote{As a word of caution, we should add that the
good agreement shown in figure \ref{fig:lattice3} may be somewhat
fortuitous since corrections can be expected for both the lattice
\cite{blaze} and the SYM results \cite{nextorder}.}

In particular, the plateau displayed for the QCD-like theories in
figure \ref{fig:lattice3} could indicate that these strongly
interacting theories are in a conformal phase where the temperature
provides the only relevant scale. Of course, as warned above and in
figure \ref{fig:compare2}, the coupling in QCD runs with the energy
scale and hence the temperature. Therefore, if we consider QCD at
temperatures much higher than $T_c$, we must find that it behaves as
a free gas of quarks and gluons because asymptotic freedom has
driven the theory to weak coupling. In contrast, being a conformal
theory, the SYM theory will remain at strong coupling for any
temperature. Therefore the similarities between QCD and SYM will
evaporate again at very high temperatures and it is only in some
range of temperatures above $T_c$ where the AdS/CFT correspondence
can be expected to provide useful insights. Hence we arrive at the
title of our article: to apply AdS/CFT techniques, we should cook
the QCD plasma or ``quark soup" but not too much!

Our current understanding of the AdS/CFT correspondence allows us to
consider gauge theories that are somewhat closer to QCD than SYM.
For example, as indicated in the comparison in figure
\ref{fig:compare2}, we can add a small number of flavours of
``quarks" in the fundamental representation to make a ${\cal N} = 2$
supersymmetric theory \cite{first}. %,first2}.
The supergravity dual of this theory involves extra D7-branes living
in $AdS$ background. These extra D-branes can be visualized as the
place where the open strings that represent the quarks can end.
However, technical issues have restricted these studies to consider
$N_f/N\ll 1$. The literature on these constructions is now quite
large and so we refer the interested reader to a recent review
\cite{flavorreview}. One can also find gravity duals to theories
that show confinement \cite{conf1, conf2}. Of course, this is all in
the 't Hooft limit and so it is not what we need to understand QCD.
Nevertheless, a lot has been learned about the geometrical
interpretation of confinement. Further, chiral symmetry breaking has
been implemented recently in the popular model of Sakai and Sugimoto
\cite{sakai}.

%In fact, there is a whole ``phenomenological" approach to AdS/QCD
%which tries to piece together a gravity theory that would best
%``fit" QCD data \cite{adsqcd}.

However, our focus here is not on developing a more elaborate
correspondence that comes closer to providing a complete description
of QCD. Rather, we are interested in studying the known
correspondence at finite temperature. In particular, as we
mentioned, we are interested in studying the fluid properties of the
SYM theory plasma and not it's detailed microscopic dynamics. In the
next section we will explain in more detail how to do finite
temperature CFT calculations using just Einstein's gravity.

  \section{AdS/CFT at Finite Temperature and the sQGP}

How do we calculate properties of the SYM fluid that we can compare
with the sQGP? First, we need to learn how to put SYM at finite
temperature. For a constant temperature, we will have a static
homogeneous plasma. The dual of this phase is given by the so-called
black three-brane background, which is simply a black hole embedded
in $AdS_5$ \cite{conf1}. The five-dimensional metric may be written
as
 \begin{equation} \label{bh}
ds^2 = \frac{(\pi T)^2 }{u} \left[ - f(u) dt^2 + d\vec{x}^2 \right]
+ \frac{du^2}{4 u^2 f(u)}\;,
 \end{equation}
where $f(u) = 1 - u^4$ and $T$ is the Hawking temperature. In these
coordinates, the horizon is located at $u = 1$ and the asymptotic
boundary at $u = 0$.

The calculation of hydrodynamic transport coefficients using this
background was pioneered by \cite{starinets2,starinets1}. A useful
review of these results and the subsequent developments appears in
\cite{starinetsreview}. In the original calculations, the transport
coefficients, such as the shear viscosity, were determined from
correlation functions of the stress tensor using the so-called Kubo
formulae. These correlation functions were calculated using the
AdS/CFT dictionary. Here we will follow a slightly different but
equivalent approach, which we think is less technical. We will
directly calculate the stress tensor and will compare it with the
expectation from hydrodynamics.

Let us do this first for the black hole background, Eq. (\ref{bh}).
We can put this metric in the Feffermam-Graham form (\ref{metric2})
by making the change of variables:  $u ={2 \rho}/({1 + \rho^2})$. We
can then expand at small $\rho$ as explained in section 2. Reading
off the asymptotic coefficients as in Eq. (\ref{asymptotic}), we can
easily calculate the stress tensor using Eq. (\ref{T}). We leave the
details for the interested reader. The result is,
 \begin{equation} \bra T_{\mu \nu}\ket_{SYM} = \frac{\pi^2 N^2 T^4}{8}
\left(
\begin{array}{cccc}
 3 &  0 & 0 & 0 \\
 0 & 1  &  0&0 \\
  0 &  0  & 1 &0\\
   0 &  0  & 0 &1
   \end{array}
\right)\;.
\end{equation}
This is precisely the expected stress tensor for a conformal fluid
at equilibrium with energy density $\varepsilon = 3\, p =
\frac{3\pi^2}{8}\, N^2\, T^4$. Here the factor $T^4$ reflects the
usual Stefan-Boltzman law while $N^2$ reflects the number of degrees
of freedom in the plasma. Further note that since the energy and
pressure densities are related as $\varepsilon=3\,p$,
$T^\mu{}_\mu=0$ as expected for a conformal field theory.

Now let us take the plasma a bit out of equilibrium. For a
(conformal) fluid out of equilibrium, we can write down the most
general stress tensor as,
 \begin{equation}
T_{\mu \nu} =  \varepsilon\, u_{\mu} u_{\nu} + p \,\Delta_{\mu \nu}
+ \Pi_{\mu \nu}\;,
 \end{equation}
where $u^\mu$ is the fluid 4-velocity and the projector $\Delta_{\mu
\nu} = g_{\mu \nu} + u_{\mu} u_{\nu}$ satisfies $\Delta_{\mu
\nu}u^\nu=0$.
%\footnote{For the reader familiar with fluid
%dynamics, we need to mention that we are using the Landau-Lifshitz
%or energy frame when writing the stress tensor.}
The shear tensor $\Pi_{\mu \nu}$ is also orthogonal to the
4-velocity, \ie $\Pi_{\mu \nu} u^\nu = 0$, as well as traceless
$\Pi^\mu{}_\mu = 0$. Moreover, it vanishes at equilibrium.

Now, for a general fluid, the conservation equation
\begin{equation} \nabla_\mu T^{\mu \nu} = 0\;,\end{equation}
only determines the energy density and four-velocity of the fluid.
The shear tensor remains undetermined and it is the goal of fluid
theory to provide extra equations to determine this quantity. One
usually writes such equation as an expansion in derivatives of
hydrodynamical quantities. To lowest order, we have
 \begin{equation}\label{Pi}
\Pi_{\mu \nu} = -2\, \eta\, \sigma_{\mu \nu} + \ldots\;,
 \end{equation}
where the constant $\eta$ is the shear-viscosity and
 \begin{equation}
\sigma_{\mu \nu} = \Delta_{\mu}^\alpha \Delta_\nu^\beta
\,\nabla_{(\mu} u_{\nu)} - \frac{1}{3} \Delta_{\mu \nu}
\nabla_\alpha u^\alpha\;.
 \end{equation}

Given this framework from fluid mechanics, our goal now is to
calculate the shear viscosity using the AdS/CFT correspondence. To
perturb the fluid, we create a linearized metric perturbation in the
bulk. That is, we will solve the linearized Einstein's equations in
five dimensions (with a negative cosmological constant) for a metric
perturbation of the form: $\delta g_{x y}  := (\pi T)^2 h_{x y}(u,
x)/u $. This bulk perturbation will also induce a perturbation in
the metric on the boundary given by $h_{x y}(0,x) := h_{x y}(x)$.

Now, we have written the fluid equations above for an arbitrary
metric. Therefore, we can expand them around the static plasma in
flat space. It is an easy exercise to show, using Eq.~(\ref{Pi}),
that the shear tensor is modified, to linear order in the metric
fluctuation, by $ \Pi_{x y} = - \eta\, \partial_t h_{x y}(x) +
\ldots$, where the dots denote higher derivative terms. That is, we
are using a metric fluctuation in the boundary to ``shake" the
fluid.

In practice, one goes to Fourier space and writes
 \begin{equation} \label{Pigauge}
\tilde \Pi_{x y} (\omega, \vec k ) =  \tilde h_{x y}(\omega, \vec{
k}) \left[ i\, \eta \,  \omega  + {\cal O}(\omega^2
,\vec{k}^2)\right]
 \end{equation}
We can do the same in the bulk and solve Einstein's equation for
each Fourier component $\tilde h_{xy}(\omega, \vec{k}; u)$ with
$\tilde h_{xy}(\omega, \vec{k}; u = 0) = \tilde h_{x y}(\omega,
\vec{ k}) $. When doing this, one needs to impose some boundary
conditions at the horizon $u = 1$. These were discussed in
\cite{starinets1,starinets5}. The first is regularity at the horizon
(ensuring that we stay within the linearized approximation). The
second is that each Fourier mode should have the form of an {\it
incoming} wave towards the horizon. That is, we do not want any
fluctuations coming out of the horizon.  In \cite{starinets1}, this
last condition was related to a choice of {\it retarded} Green's
function for the gauge theory.

To lowest order in the frequency $\omega$ and setting $\vec{k} = 0$
for simplicity, one can easily show that the solution has the form:
 \begin{equation}
\tilde h_{x y}(\omega,u) = \tilde h_{xy}(\omega) (1- u)^{-i
\omega/(2 \pi T)}\left[1  + \frac{i\, \omega }{4 \pi T} \log(1 + u)
+ {\cal O}(\omega^2)\right] \;.
 \end{equation}
Expanding the full metric (with the fluctuation) around the
asymptotic boundary, we can read off the response of the stress
tensor using  Eq. (\ref{T}). We get,
 \begin{equation} \label{Pigrav}
\tilde \Pi_{x y}(\omega)= \tilde h_{x y}(\omega)\left[  i\,  \omega
\left(\frac{ \pi N^2 T^3}{8}\right) +{\cal O}(\omega^2)\right]
 \end{equation}
Hence comparing the results (\ref{Pigauge}) and (\ref{Pigrav}) we
find the value of the shear viscosity of the ${\cal N} = 4$ SYM
plasma to be
\begin{equation} \eta = \frac{\pi}{8}\,N^2 \,T^3 \;.\end{equation}

Another important result derived from this calculation is the value
of the ratio of the viscosity to the entropy density. Using the
thermodynamic relation $s =  4 \varepsilon/(3T)$ (at zero chemical
potential), we have
 \begin{equation}\label{eq}
\eta/s = 1/4 \pi\;.
 \end{equation}
This last result has been the focus of much attention in the past
years. It turns out that one can prove that the equality (\ref{eq})
is {\it universal} for all gauge theories with a gravity dual, in
the limit where the gravity approximation is valid (which is usually
in the 't Hooft limit) \cite{universality1, universality2}. In fact,
it has been argued that {\it any} fluid that can be obtained from a
quantum field theory, will obey the bound \cite{universality1}:
 \begin{equation}\label{ineq}
\eta/s \geq 1/4 \pi\;.
 \end{equation}
This is indeed a very strong statement. Some support for this bound
comes from calculations which indicate that the leading corrections
in $1/\lambda$ to the SYM result \reef{eq} in fact increase the
ratio so that the bound is no longer saturated \cite{nextorder}.
Certainly in nature, no laboratory tests to date have found a
substance which violates this bound. However, this conjectured bound
\reef{ineq} has drawn particular attention by preliminary
investigations of the experimental data from RHIC which indicated
that the sQGP had an unusually low viscosity \cite{teaney}.
Determining a precise value of the viscosity in the sQGP is now an
topic of intense study but recent investigations seem to indicate
that $\eta/s\sim1/4\pi$  --- for example, see \cite{intent}.

%\subsection{Recent Developments}

These unexpected results for the viscosity combined with a lack of
alternative theoretical tools to study the strongly coupled dynamics
of the sQGP have stimulated tremendous activity in calculating
different thermal properties of strongly coupled non-abelian plasmas
for the ${\cal N} = 4$ SYM and other holographic theories. We are
not able to cover all of the literature on this subject but in the
following we point the interested reader towards some of the
interesting developments.

Above, we showed how to compute the shear viscosity but this only
determines the lowest derivative component of the shear tensor
$\Pi_{\mu \nu}$, as shown in Eq.~\reef{Pi}. In principle, there are
an infinite series of higher order terms, each one characterized by
a different transport coefficient. At present, there is no complete
theory of relativistic fluid dynamics that predicts the form of all
such terms.  The ``standard" theories are of second order in
derivatives \cite{hydrotheories}. In principle one can extend the
holographic calculations to higher orders in frequencies and
wave-vectors, to uncover the rest of the terms. This procedure has
recently been carried out to second order in derivatives in
\cite{starinets6}. Identifying the correct set of terms will be
crucially important in refining the analysis of the experimental
data to determine the precise value of $\eta/s$ for the sQGP. There
have also been calculations of transport coefficients in
non-conformal field theories with gravity duals \cite{alexN2}.  Of
course there have also been a variety of other holographic
calculations including: studying the effects on introducing a finite
chemical potential \cite{mu}, investigating spectral functions
\cite{spectre}, examining the diffusion of heavy quarks
\cite{diffuse} and calculating the ``photon" emission rate of the
plasma \cite{emission}.
%\footnote{Note that SYM theory has no
%photons. However, one can imagine coupling the theory to a $U(1)$
%gauge theory. To lowest order in the electric coupling, one can
%relate the photon production rate to a two-point function of the SYM
%theory only.}

To conclude then, theorists face many new challenges in developing a
physical understanding of the recently discovered strongly coupled
quark-gluon plasma. In parallel developments, however, string
theorists have developed the AdS/CFT correspondence as a new
analytic tool to study certain gauge theories, \eg ${\cal N}=4$ SYM,
which are well suited for this purpose. In particular, in the
strongly coupled regime, these gauge theories are dual to a
(super)gravity theory in higher dimensions. Further, although these
gauge theories differ from QCD in many details, at finite
temperature, they seem to share many features in common with the
sQGP. Hence gravitational calculations are being used to gain
insight into this new phase of QCD as the implications of this
remarkable correspondence continue to be explored. At the same time,
experimentalists will soon begin to explore a new frontier with
heavy ion collisions at $\sim$ 5 TeV/nucleon at the Large Hadron
Collider
--- see figure \ref{fig:lattice3}. Thus we can expect to see
new surprises coming from both theory and experiment in the near
future.

\ack This article provides an informal summary of a talk given by
RCM at the 18th International Conference on General Relativity and
Gravitation in Sydney, Australia, July 8 -- 13, 2007. RCM would like
to thank the organizers of {\it GRG18} for the opportunity to speak
at such an engaging meeting in such a pleasant setting. Research at
Perimeter Institute is supported by the Government of Canada through
Industry Canada and by the Province of Ontario through the Ministry
of Research \& Innovation. RCM also acknowledges support from an
NSERC Discovery grant and funding from the Canadian Institute for
Advanced Research.

\end{document}